\documentclass[conference]{IEEEtran}
\usepackage{cite}
\usepackage{amsmath}
\usepackage{graphicx}
\usepackage{lipsum}
\usepackage{tikz}
\usepackage{url}

\begin{document}

\title{Hypothetical Problems concerning the Theory of Relativity on Cryptographic Currency Implementations}
\author
{\IEEEauthorblockN{Abrahim Ladha}
\IEEEauthorblockA{Computer Security 5700\\
Armstrong State University\\
abrahimladha@protonmail.ch}
}
\maketitle

\begin{abstract}
Bitcoin has demonstrated there are many security improvements applicable to normal currency. As the human race expands and we colonize other planets, we have to consider how we are going to extend integral parts of of society, and that includes our currency system. Information transferring does not scale well with very large distances, entirely due to physical limitations. For example, there is a maximum speed that any information can travel, and it cannot be faster than the speed of light. In this paper we take these physical limitations into account to give treatment to the following question. Can a single crypto-currency be used across the entire universe? Trivially many currencies can be used with exchange rates but we will try to avoid this as our solution. The idea of this paper was inspired by a paper titled "The Theory of Interstellar Trade" by Paul Krugman, a Nobel Economist. \cite{krugman}
\end{abstract}

\section{What is bitcoin?}
\textbf{History} Bitcoin was invented by someone using the pseudonym Satoshi Nakamoto. In his original whitepaper  \cite{nakamoto} titled "Bitcoin: A Peer-to-Peer Electronic Cash System", he introduces the concept of a decentralized and cryptographically secured monetary system. He also covers topics on proof-of-work, transactions, mining, privacy, and attacks on the network. It was implemented as free software and released in January 2009. Unlike gold, bitcoin has no attachment to any sort of industry, so the price day to day price fluctuation is really based on nothing but speculation. The US Dollar technically isn't tied to any sort of industry, but it is backed and regulated by the Fed. In contrast, bitcoin has no central authority and fluctuates relatively wildly. It has been as high as 1000 USD per bitcoin.\cite{1000usd}

\textbf{Units} There can only ever exist a maximum of 21 million bitcoins. Each bitcoin can be broken up into a hundred million pieces, much like a single US Dollar can be broken up into one hundred pennies. The smallest unit of bitcoin is called a satoshi. $10^8$ satoshis equal one whole bitcoin. \cite{units}

\textbf{Ownership} Since bitcoin has no central authority, transactions are made peer to peer. A user can only send bitcoins to another user if they can digitally sign the transaction with their private key. Without the private key, a malicious user cannot sign the transaction and the coins cannot be spent. \cite{ownership}

\textbf{Transactions} A transaction is a data structure with a source of funds for the input and a destination, or the output. A bitcoin transaction is just 300 to 400 bytes of data. Once a bitcoin transaction is sent, it will be validated by that node. If valid it will propagate through the nodes to which it is connected sending a confirmation to the sender. The propogation grows exponentially across the network until everyone has received the message. To prevent spamming and denial of service attacks, each node independently validates each transaction before propagating it further. A bitcoin transaction has what is called \textit{unspent transaction output}, or UTXO. These are indivisible units of bitcoin with an associated owner. These are recognized as currency units by the rest of the network and recorded into the blockchain. an owners bitcoin amount would be scattered UTXO from many transactions and many blocks. \cite{trans}

\begin{figure}[ht!]
\includegraphics[scale=.3]{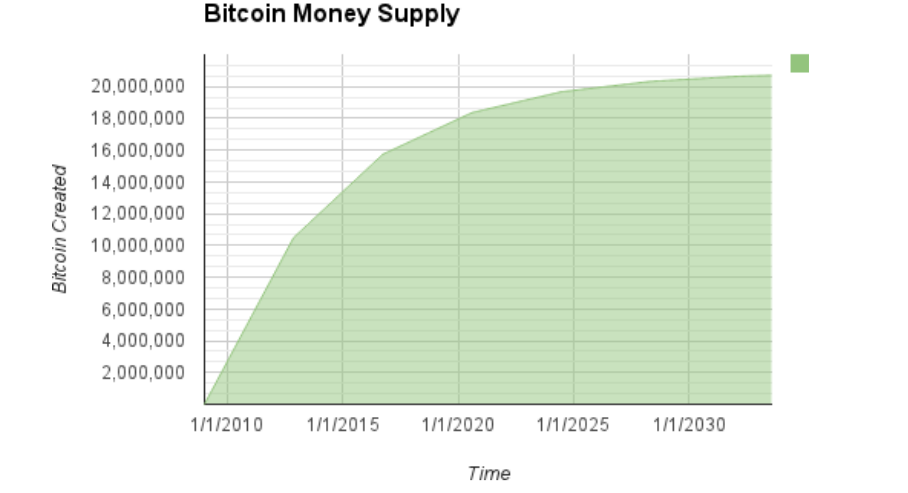} 
\caption{Bitcoins mined vs time. Notice the rate halvening every so often so that bitcoins add at a diminishing rate.}
\end{figure}

\textbf{mining} In bitcoin, trust is based upon computation. Transactions are grouped into blocks which are NP hard to solve (very difficult to prove, but very easy to verify as proven). Mining has two main purposes. Mining creates new bitcoins with each block which serves the role that the Federal Reserve and Mint play for the US Dollar. The bitcoin reward per block is fixed for a certain amount of time until a "halvening" occurs on some predetermined events. The bitcoin reward per block continues to half so that coins are added at a diminishing rate. Mining is also the heart of verifying transactions. Mining also creates trust by confirming transactions only if enough computation power was dedicated to the block containing the transaction. The actual action of mining itself is simply trying to find a solution to the proof-of-work algorithm which verifies the block. The hash function used in bitcoins mining process is SHA256. Bitcoin's proof-of-work algorithm is very basic. It simply guesses a string, hashes it, and if it doesn't match the correct hash, it increments its nonce. The miner constructs a candidate block and fills it with transactions. It then calculates the hash of the block header and compares it to the current target. If its bigger, it’ll increment the nonce and and try again. Today, miners do this quadrillions of times per block. \cite{hashcash}

Blocks contain a parameter known as difficulty bits. From this the difficulty target can be calculated. The formula for calculating the difficulty is: 
\begin{equation}
target = C_02^{8(e_o - 3)}
\end{equation}

For example, in block 277316, the difficulty bits is $0x1903a30c$. The coeffcient $C_0 = 0x19$ and the exponent $e_o = 03a30c$ In decimal, the difficulty target is roughly $2.2 \times 10^{58}$ which is huge. \cite{bitcoincore}

\section{The Theory of Relativity}
\textbf{history} In physics up until the point of these discoveries, there had been major inconsistencies with the laws of physics between mechanics and electromagnetism. In 1905, Albert Einstein published a series of papers as part of his \textit{Annus Mirabilis} showing that just with two simple postulates, classical physics was simply an approximation that got more inaccurate at very large values. The Theory of Relativity was the puzzle piece to fix these inconsistencies. Consequences of Relativity are even proved to this day, with the recent discovery of gravitational waves. \cite{history,waves}

\textbf{Postulates} Einstein's Theory of Relativity is based upon two postulates: 
\begin{enumerate}
\item The Laws of Physics are the same for all observers in all inertial reference frames. No one frame is preferred over the other.
\item The speed of light in a vacuum has the same value $c$ in all directions and in all inertial reference frames.
\end{enumerate} 

Galileo's postulates stated that the laws of \textit{mechanics} were the same in all inertial frames. Einstein extended this to all laws of physics, to include optics, and electromagnetism. The first postulate does not say that all quantities are measured the same for all inert observers, but the laws, the relationship between quantities, is the same. 

The second postulate is more defining. Previously we had thought that there was no speed limit to anything, and you simply required enough energy to reach that speed then it was attainable. We now realize that no information can travel faster than light. We denote the speed of light as the constant $c = 299792458$m/s. No particle with mass can even reach this speed. \cite{einstein}
\begin{figure}[ht!]
\includegraphics[scale=1]{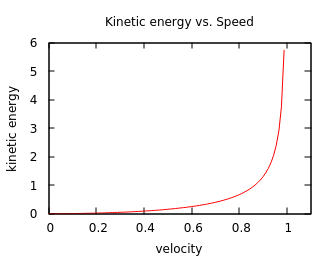} 
\caption{Kinetic energy required to reach a certain velocity as a fraction of the speed of light. If you have mass, to reach the speed of light would require infinite energy.}
\end{figure}

\textbf{Lorentz Transforms} The Lorentz transforms are a system of equations that can be derived from Einstein's Postulates. They are for transforming reference frames. We state them without proof:
\begin{equation}
x' = \gamma(x - vt)
\end{equation}
\begin{equation} 
t' = \gamma(t - vx/c^2)
\end{equation}
\begin{equation}
\gamma = \frac{1}{\sqrt{1 - (v/c)^2}}
\end{equation}
Notice that $t'$ is dependent on position, that is to say, that space and time are entangled. This was a fundamental principle of Einstein's theory, one that was long rejected by his contemporaries. Why hadn't these been derived before? Well let $c \rightarrow \infty$ and our equations become $x' = x - vt$ and $t' = t$. These are the classical Galilean transforms, which work just fine at small speeds compared to $c$. From these, one could have deduced (incorrectly) that time passes at the same rate for all frames of reference. \cite{lorentz}

Suppose two events occur at the same place in some reference frame, but at different times, then (3) reduces to:
\begin{equation}
\Delta t = \gamma \Delta t'
\end{equation}
What this means is the faster you go relative to some frame, the slower time gets for you relative to that frame. For example, if you leave earth when you are born on a rocket going 0.99$c$ relative to earth, When the earth clock says it should be your 100th birthday, the rocket clock will say that you are only a little over 14 years old, and you will look and feel only 14. 

If a rod is at rest in some reference frame, then any observer in that frame and easily measure its length by subtracting the positions of its two endpoints, that is to say, $L = \Delta x$. Suppose the rod is moving. The length of the rod can only be measured if the endpoints are measured \textit{simultaneously}, which is to say $\Delta t = 0$, then (2) reduces to: 
\begin{equation}
\Delta x = \frac{\Delta x'}{\gamma}
\end{equation}
What this means is that the faster something is moving relative to you, the longer it will appear. For example lets say you are in the same rocket going 0.99$c$. If you on your rocket measure it end to end to be 100 meters, then on earth they will measure it to be a little more than 708 meters. This effect is often portrayed in sci-fi when a spaceship "spaghettifies" as it enters hyperdrive or a wormhole or whatever. \cite{spaghetti}

\textbf{Example Problem} To demonstrate some of this material, we present a famous example problem. There are two planets known to be hostile towards one another, $A$ and $B$ which are $4.0 \times 10^8 $m apart. You are in a rocket traveling at $0.98c$ relative to $A$ and $B$. Your rocket follows a straight line, first past $A$, then $B$. You detect a high energy microwave signal from $A$ and then, 1.10s later, an explosion on planet $B$. Clearly $A$ has attacked $B$. Should you prepare for a confrontation? First we set up our reference frame. We let the rocket be stationary and the $A$-$B$ planetary system moving at $0.98c$ relative to the rocket. We could have chosen another equivalent reference frame but this will make calculations simpler. let $x_A$ and $x_B$ denote the position of the signal from $A$ and the explosion on $B$ respectively, and $t_A$ and $t_B$ the times. Therefore $\Delta x = x_B - x_A = +4.0 \times 10^8$m, and $\Delta t = t_B - t_A = +1.10$s. We now transform the reference frame to that of $A$-$B$ and calculate $\Delta t'$ and $\Delta x'$. with $v = 0.98c$, $\gamma = 1/\sqrt{1-(v/c)^2} = 1/\sqrt{1-(0.98c/c)^2} = 5.0252$. Therefore $\Delta x' = \gamma(\Delta x - v\Delta t) = 3.86 \times 10^8$m, and $\Delta t' = \gamma (\Delta t - v\Delta x/c^2) = -1.04$s. Well $\Delta t'$ is negative. What does this mean? Well $\Delta t' = t'_B - t'_A = -1.04$ seconds. This tells us that $t'_A > t'_B$ which implies that the signal happened 1.04 seconds \textit{after} the explosion. But we witnessed the signal \textit{before} the explosion, so which is it? If there is a relationship between these events, then information must travel from one to the other. If we check the speed of this information, we see $v_{info} = 4.0 \times 10^8$ meters / $1.10$ seconds $ = 3.64 \times 10^8$ m/s. But this speed is impossible since it exceeds $c$. Therefore neither event is dependent on the other, and these are unrelated events.
 
\section{CHOOSING OUR BLOCKTIME}
The greatest constraint on mining in space is how long it takes information to travel huge distances. If our blocktime is 10 minutes, and it takes light 11 minutes to travel to a planet with a dominating amount of computing power, the probability of shares of worked submitted being valid is a statistical improbability. We however, when designing a currency system can increase or decrease the blocktime. In the next several sections, we give example scenarios, and structures to determine a blocktime. 

In the current bitcoin implementation, the blocktime is a fixed 10 minutes. This was chosen by Satoshi as a trade-off between the first confirmation time and the amount of work wasted due to chain splits. Shorter block times have the benefit of a faster first confirmation time (greatly reducing the probability of double spending) and less pay out variance for miners. Smaller blocktimes have downsides as well, like more bandwidth, more and longer forks, longer reorganization time and a greater portion of the raw hashpower is wasted, lowering security. These are worthy trade-offs for the fast first confirmation time. 

\section{GRAPH THEORETIC MODEL OF OUR NETWORK}
We first contruct the structure in which we will base our network models on. Our structure will be a simple graph with weighted edges. Each node will represent a user, a miner, or a group of miners on our network. An edge will form between two nodes if and only if they are connected peers on the network. Not all peers connect to each other so our graph is not necessarily complete. The weight of an edge is a variable real number which represents the time in which it takes information travelling the speed of light to reach the other node. We will consider nodes travelling with non zero velocity but none of the cases of acceleration. This puts us squarely in the realm of special relativity. 
 
\textbf{pool mining of two nodes} Consider the of only two simplified nodes. For this example we use the Earth-Mars planetary system. All the miners on Earth have formed a pool, as have the miners on Mars. transactions from Earth to Earth or Mars to Mars is no big deal, and would work similarly to how transactions work today. If a transaction from Earth to Mars is created, sending the information at the speed of light would take over seven minutes just to reach the other planet. 
\begin{tikzpicture}[auto]

\node[draw,shape = circle] (earth) {Earth};
\node[draw, shape = circle](mars) at (6,2) {Mars};

\node[draw, shape = circle](mars1) at (6,3) {};
\node[draw, shape = circle](mars2) at (6.7,2.7) {};
\node[draw, shape = circle](mars3) at (7,2) {};

\node[draw,shape = circle] (earth1) at (0,-1) {};
\node[draw,shape = circle] (earth2) at (-0.7,-.7) {};
\node[draw,shape = circle] (earth3) at (-1,0) {};

\path (earth1) edge node {} (earth);
\path (earth2) edge node {} (earth);
\path (earth3) edge node {} (earth);

\path (mars1) edge node {} (mars);
\path (mars2) edge node {} (mars);
\path (mars3) edge node {} (mars);

\path (mars) edge node {7.533 minutes} (earth);

\end{tikzpicture}

If we choose planets farther and farther apart for our example, the distance between them would become larger and larger and sending shares would take more time. If a transaction is sent, the entire network must also verify the transaction before a new one can proceed. We will discuss this more in the section on propagation.
Because we only have two main nodes, we only have two cases. If one is greater than the other or if they are both equal. The second case is a statistical improbability so we will not consider it (All it takes is one user on either side to drop out of the pool for this case to reduce back to the first case). Consider the case where Earth is greater than Mars. Since there are only two nodes, Earth must have greater than 50\% of the network. Then nearly 100\% of the blocked solved would be on Earth, as if a miner on Mars wants to submit shares, The delay between Earth and Mars must be accounted for, and as that distance increases, the probability that shares from some planet arrive stale to Earth increases as the distance increases. The 51\% attack is also extremely likely, and Earth miners could censor transactions to and from Mars.\\
\textbf{pool mining of n nodes} Now we consider generalizing the previous case to some $n$ nodes. 

\begin{tikzpicture}[auto]

\node[draw,shape = circle] (earth) {Earth};
\node[draw,shape = circle](mars) at (6,2) {Mars};
\node[draw, shape = circle](venus) at (2,4) {Venus};

\node[draw, shape = circle](mars1) at (6,3) {};
\node[draw, shape = circle](mars2) at (6.7,2.7) {};
\node[draw, shape = circle](mars3) at (7,2) {};

\node[draw,shape = circle] (earth1) at (0,-1) {};
\node[draw,shape = circle] (earth2) at (-0.7,-.7) {};
\node[draw,shape = circle] (earth3) at (-1,0) {};

\node[draw,shape = circle] (venus1) at (3,4) {};
\node[draw,shape = circle] (venus2) at (2.7,4.7) {};
\node[draw,shape = circle] (venus3) at (2,5) {};

\path (earth1) edge node {} (earth);
\path (earth2) edge node {} (earth);
\path (earth3) edge node {} (earth);

\path (mars1) edge node {} (mars);
\path (mars2) edge node {} (mars);
\path (mars3) edge node {} (mars);

\path (venus1) edge node {} (venus);
\path (venus2) edge node {} (venus);
\path (venus3) edge node {} (venus);

\path (mars) edge node {$\alpha$} (earth);
\path (venus) edge node {$\beta$} (earth);
\path (mars) edge node {$\delta$} (venus);

\end{tikzpicture}

With $n > 2$ or more nodes, We avoid the immediate outcome of the 51\% attack. For a transaction to be sent from say, Earth to Mars, The transaction would be confirmed by only Earth and Mars if $\alpha > \beta$, but bitcoin doesn't quite work that way. UTXO would be scattered all over these $n$ planets, requiring their participation in a transaction. As a worst case scenario, If we let $\alpha < \beta < \delta$ and $\alpha + \beta < \delta$ then the  time to confirm a transaction would $2(\alpha + \beta)$.

\textbf{Lattice} Consider the case of evenly space miners in a lattice in three dimensional euclidean space. 
\begin{center}
\begin{tikzpicture}
\def \dx{1.66};
\def \dy{2.4};
\def \dz{1.66};
\def \nbx{4};
\def \nby{4};
\def \nbz{4};
\foreach \x in {1,...,\nbx} {
    \foreach \z in {1,...,\nbz}{
        \draw (\x*\dx,\dy,\z*\dz) -- ( \x*\dx,\nby*\dy,\z*\dz);
    }
}
\foreach \y in {1,...,\nbx} {
    \foreach \z in {1,...,\nbz}{
        \draw (\dx,\y*\dy,\z*\dz) -- ( \nbx*\dx,\y*\dy,\z*\dz);
    }
}
\foreach \x in {1,...,\nbx} {
    \foreach \y in {1,...,\nbz}{
        \draw (\x*\dx,\y*\dy,\dz) -- ( \x*\dx,\y*\dy,\nbz*\dz);
    }
}
\foreach \x in {1,...,\nbx} {
    \foreach \y in {1,...,\nby} {
        \foreach \z in {1,...,\nbz} {
            \node at (\x*\dx,\y*\dy,\z*\dz) [draw, circle, fill=white] {};
        }
    }
}
\end{tikzpicture}
\end{center}
Our worst case scenario for a transaction is that there exists UTXO on every single node, and the two nodes participating in a transaction are on opposite corners. Let all the edges be the same length and the lattice of length $l$, width $w$, and height $h$. we can represent the weight of each edge as $\alpha$. Since we don't have any sort of diagonals, all possible paths that only go left, down, and back (as in, they don't go backwards or in loops) are of equal length. For our propagation, we have to consider every shortest path from the starting node. Since every node is on a shortest path, and every shortest path is equal, then every node will be propagated to the end node at the same time. That time would be $\alpha \times lwh$.

\section{rotational scenarios}
\textbf{planet - satellite relationship}
Consider the case of a planet and satellite relationship.

\begin{center}
\begin{tikzpicture}[auto]

\node[dashed,draw,shape = circle] (orbit1) at (0,0) {~~~~~~~~~~~~~~~~~~~~~~~~~~~~~~~~~~~~~~~~~~~~~~~~};
\node[fill=white,draw,shape = circle](p2) at (3,0) {$p_2$};

\node[fill = white, draw,shape = circle] (p1) at (0,0) {$p_1$};

\path[->] (p1) edge node {$r_1$} (p2);

\end{tikzpicture}
\end{center}
We wish to derive a minimum blocktime for these two entities to communicate fairly. If we henceforth denote $b$ as our blocktime, it should be obvious we choose our bloctime to be $b > \frac{r_1}{2}$

\textbf{concentric orbits}
Consider the case of two planetary bodies in the same solar system. They orbit some large mass at the center, like a sun or a black hole. 
\begin{center}
\begin{tikzpicture}[auto]

\node[dashed,draw,shape = circle] (orbit1) at (0,0) {~~~~~~~~~~~~~~~~~~~~~~~~~~~~~~~~~~~~~~~~~~~~~~~~};
\node[fill=white,draw,shape = circle](mars) at (3,0) {$p_2$};

\node[dashed, draw,shape = circle] (orbit2) {~~~~~~~~~~~~~~~~~~~~~~~~~~~~~~~~};
\node[fill = white,draw,shape = circle](earth) at (-2,0) {$p_1$};

\node[fill = black, draw,shape = circle] (center) at (0,0) {};

\path[->] (center) edge node {$r_1$} (earth);
\path[->] (center) edge node {$r_2$} (mars);

\end{tikzpicture}
\end{center}

We want to choose a blocktime in which these two planets can still submit shares to each other reasonably. Well The maximum distance these two planets can ever be apart is the sum of the two radii. The blocktime should then be chosen to be greater than this worst case scenario. The blocktime should be greater than the average of these two radii, that is: $b > \frac{r_1+r_2}{2}$. To generalize this for some $n$ planets, simply pick $p_1$ to be the planet of smallest radius and $p_2$ to be the planet of largest radius.

\textbf{separate orbits}
Consider the case of planets in two separate systems.
\begin{center}
\begin{tikzpicture}[auto]

\node[dashed,draw,shape = circle] (orbit1) at (0,0) {~~~~~~~~~~~~~~~~~~~~~~~~~~~~~~~~~~~~~~~~~~~~~~~~};
\node[draw, shape = circle, fill= black](p1) at (0,0) {};

\node[dashed,draw,shape = circle] (orbit2) at (3.4,0) {~~~~~~~~~~~~~~~~~~~~~~~~~~~~~~~};
\node[draw, shape = circle, fill= black](p2) at (3.6,0) {};
\node[fill=white,draw,shape = circle] (f1) at (-3,0){$p_1$};
\node[fill=white,draw,shape = circle] (f2) at (5.3,0){$p_2$};

\path[<->] (p1) edge node {$\alpha$} (p2);
\path[->] (p1) edge node {$r_1$} (f1);
\path[->] (p2) edge node {$r_2$} (f2);

\end{tikzpicture}
\end{center}

This ends up being a very similar case. We choose our blocktime here to be $b > \frac{r_1 + \alpha + r_2}{2}$. This generalizes the same as would the scenario of pool mining on $n$ nodes (section IV).

\section{Propogation}

\textbf{blockchain sync problems}
In our past examples, as our networks grow bigger and bigger, they become more and more infeasible. Consider the case where you have UTXO stuck in a part of the universe that takes millennia to travel to. Will your transaction ever even process? It makes much more sense to have small contained networks of distance currencies to limit the spread of UTXO, and then to simply have an exchange rate between networks. This answers our final question: Can a single currency be used intergalatically? The answer depends on the size of our network, but in general, the answer is no. It will work much more efficiently if we just have separate individual currencies, and exchange rates between them.
As the networks grows and grows substantially, the probability of a fork occurring becomes greater and greater, as it takes more and more time for a transaction to confirm before another one can occur. Orphan chains would grow longer and longer. Our blocktime would have to be chosen from the longest shortest path (the graph diameter), and  absolutely nothing we can do to prevent that. We have a completely unavoidable physical constraint on our transfer of information. As distance increases and our network expands, all our times increase, including times for propagation, mining, and confirming transactions. To conclude: A universal decentralized currency is infeasible. 

\bibliographystyle{IEEEtran}

\begin{thebibliography}{9}
\bibitem{krugman}
	Paul Krugman.
	(1978).	
	\textit{The Theory of Intestellar Trade}
	(Economic Inquiry)
	[Article].
	Available: \url{http://www.princeton.edu/~pkrugman/interstellar.pdf}
\bibitem{nakamoto}
	N. Satoshi. 
	(2008)
	\textit{Bitcoin: A Peer-to-Peer Electronic Cash System}
	[Whitepaper]
	Available: https://bitcoin.org/bitcoin.pdf
\bibitem{1000usd}
	M. Farrell. 
	(2013).	
	\textit{What Bubble? Bitcoin Tops \$1000}
	[Online].
	Available: https://blogs.wsj.com/moneybeat/2013/11/27/what-bubble-bitcoin-tops-1000/
\bibitem{units}
	The Bitcoin Wiki.
	(2015).	
	\textit{Satoshi (unit)}
	[wiki].
	Available: https://en.bitcoin.it/wiki/Satoshi\_(unit)
\bibitem{ownership}
	Jerry Brito and Andrea Castillo.
	(2013).	
	\textit{Bitcoin: a Primer for Policymakers}
	(PDF)
	[book].
	Available: \url{http://mercatus.org/sites/default/files/Brito_BitcoinPrimer.pdf}		
\bibitem{trans}
	Andreas M. Antonopoulos.
	(2014).
	\textit{Mastering Bitcoin. Unlocking Digital Crypto-Currencies. Chapter 5}
	(first edition)
	[book].
	Available: http://chimera.labs.oreilly.com/books/1234000001802/ch05.html
\bibitem{hashcash}
	Back, Adam. 
	(1997).	
	\textit{HashCash}
	[website].
	Available: http://hashcash.org/
\bibitem{bitcoincore} 
	Satoshi Nakamoto, Gavin Andresen, et al
	(2015).	
	\textit{bitcoin-core source code 0.12 main.cpp: line 675: nTargetTimespan}
	SHA256sum: 0f1cda66c841a548a07cc37e80b0727354b1236d9f374c7d44362acdb85eb3e1
	[source code].
	Available: https://bitcoin.org/bin/bitcoin-core-0.12.0/bitcoin-0.12.0.tar.gz
\bibitem{history}
	Darrigol, Olivier.
	(2005).	
	\textit{The Genesis of the Theory of Relativity}
	(Séminaire Poincaré 1: 1–22, doi:10.1007/3-7643-7436-5\_1)
	[book].
	Available: \url{http://www.bourbaphy.fr/darrigol2.pdf}
\bibitem{waves}
	B. P. Abbott et al. 
	(2016).	
	\textit{Observation of Gravitational Waves from a Binary Black Hole Merger}
	(PRL 116,061102)
	[Physical Review Letters].
	Available: \url{https://dcc.ligo.org/public/0122/P150914/014/LIGO-P150914\%3ADetection_of_GW150914.pdf}
\bibitem{einstein}
	Albert Einstein. 
	(1905).	
	\textit{On the electrodynamics of moving bodies}
	(Annalen der Physik)
	[Journal].
	Available: \url{http://fisica.ufpr.br/mossanek/etc/specialrelativity.pdf}
\bibitem{lorentz}
	Lorentz, Hendrik Antoon
	(1904).	
	\textit{Electromagnetic phenomena in a system moving with any velocity smaller than that of light}
	(Proceedings of the Royal Netherlands Academy of Arts and Sciences 6: 809–831)
	[Journal].
	Available: \url{https://en.wikisource.org/wiki/File:Electromagnetic_phenomena.djvu}
\bibitem{spaghetti}
	Wheeler, J. Craig 
	(2007).	
	\textit{ Cosmic catastrophes: exploding stars, black holes, and mapping the universe}
	(2nd ed. Cambridge University Press, p. 182)
	[Journal].
	Available: \url{http://gen.lib.rus.ec/book/index.php?md5=E102A8E9468BA09B1DBAFDD27C3D26B}




\end{thebibliography}

\end{document}